\documentclass{vgtc}                

\pdfoutput=1

\ifpdf
  \pdfoutput=1\relax                   
  \usepackage{graphicx}                
  \DeclareGraphicsExtensions{.pdf,.png,.jpg,.jpeg} 
\else
  \usepackage{graphicx}                
  \DeclareGraphicsExtensions{.eps}     
\fi%

\graphicspath{{figures/}{pictures/}{images/}{./}} 

\usepackage{microtype}                 
\PassOptionsToPackage{warn}{textcomp}  
\usepackage{textcomp}                  
\usepackage{mathptmx}                  
\usepackage{times}                     
\usepackage{cite}                      
\usepackage{tabu}                      
\usepackage{booktabs}                  
\usepackage{todonotes}
\usepackage{xspace}
\usepackage{subfigure}
\usepackage{xcolor}
\usepackage[linesnumbered,ruled,vlined]{algorithm2e}
\usepackage{algorithmic}
\usepackage[bookmarks=false]{hyperref}

\definecolor{red}{HTML}{e85641}
\definecolor{blue}{HTML}{2656ad}
\definecolor{yellow}{HTML}{e6b800}

\newcommand{\red}[1]{\textcolor{red}{#1}}
\newcommand{\blue}[1]{\textcolor{blue}{#1}}
\newcommand{\yellow}[1]{\textcolor{yellow}{#1}}

\let\oldnl\nl
\newcommand{\nonl}{\renewcommand{\nl}{\let\nl\oldnl}}

\RequirePackage{array}
  {\begingroup
   \newcommand\estyle{}%
   \renewcommand\affiliation[1]%
     {\\\multicolumn{#1}{@{}c@{}}{\scriptsize\begin{tabular}[t]{@{}>{\footnotesize}c@{}}##1\end{tabular}}}   \begin{tabular}[t]{@{}*{#1}{>{\estyle}c}@{}}
  }%
  {\end{tabular}%
   \endgroup
  }

\newcommand{\system}{\textit{Sentifiers}\xspace}

\onlineid{0}

\vgtccategory{Research}
\vgtcpapertype{technique}

\vgtcinsertpkg

\title{Sentifiers: Interpreting Vague Intent Modifiers in Visual Analysis using Word Co-occurrence and Sentiment Analysis}

\author{Vidya Setlur\thanks{e-mail: vsetlur@tableau.com}\\ %
        \scriptsize Tableau Software %
\and Arathi Kumar\thanks{e-mail: akumar@tableau.com}\\ %
     \scriptsize Tableau Software. %
}

\teaser{
  \centering
\subfigure[]{ \includegraphics[width=0.49\textwidth]{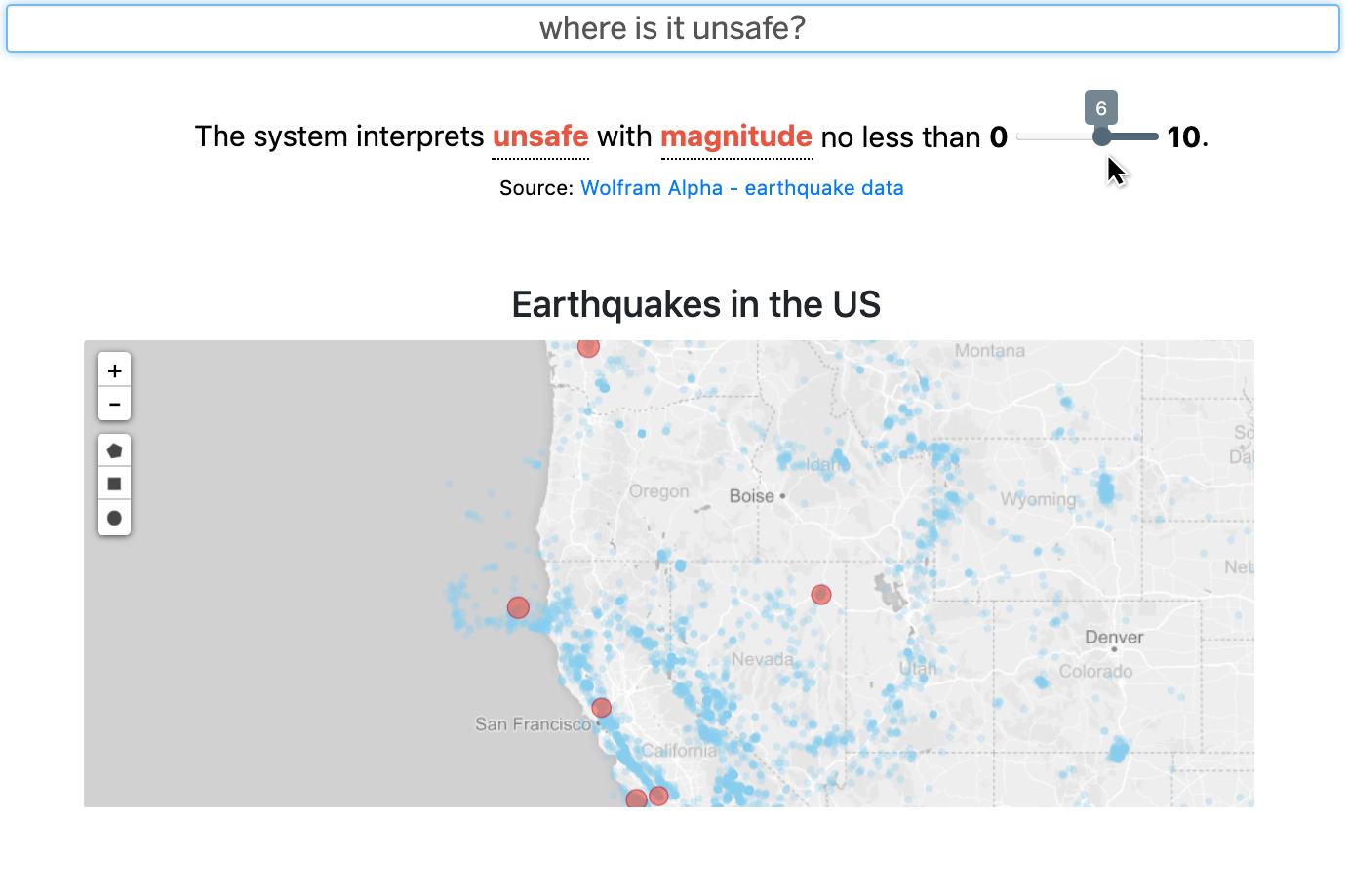}}
\subfigure[]{\includegraphics[width=0.49\textwidth]{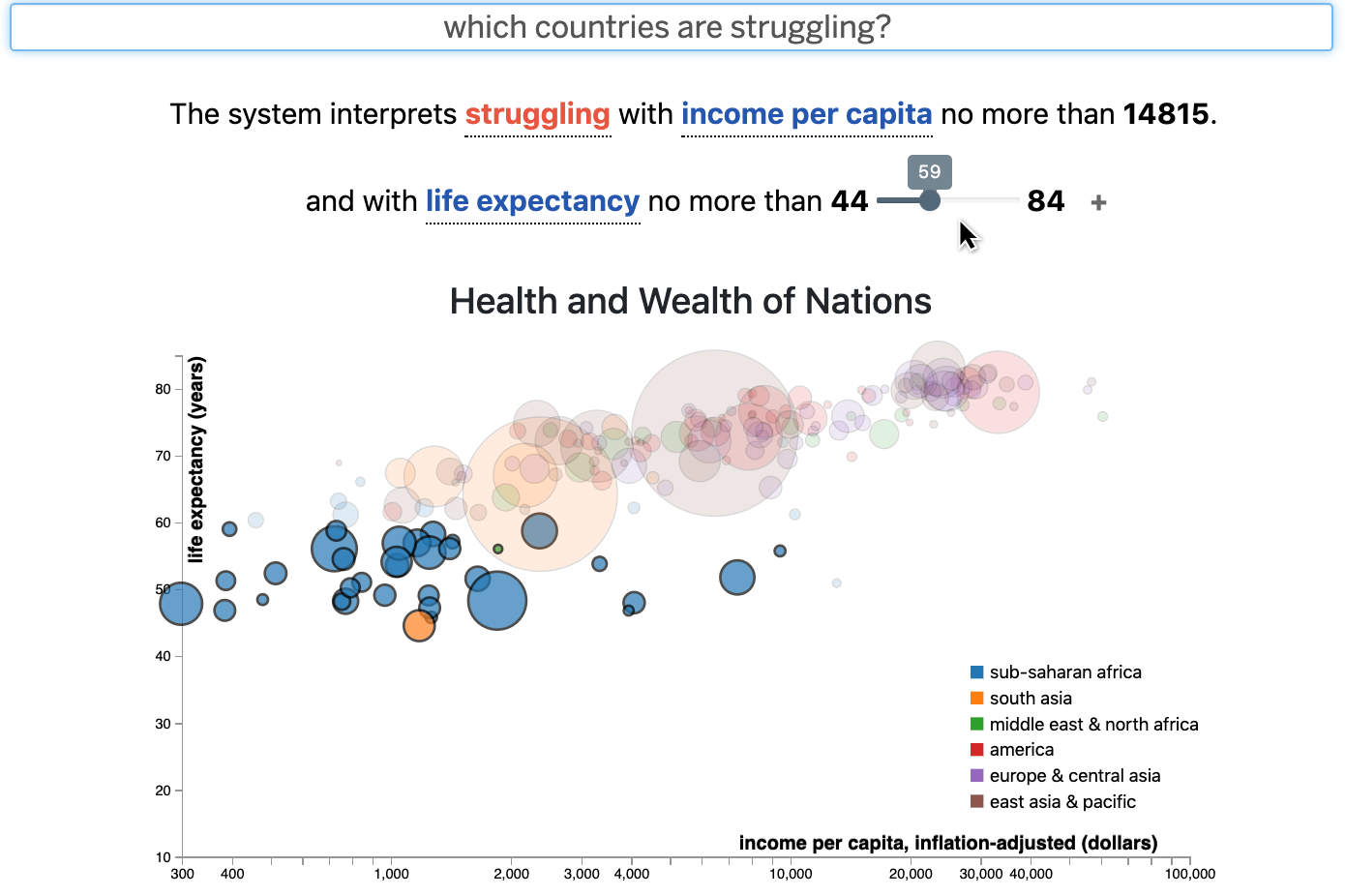}}
 \caption{Screenshots of \system showing the interpretation of vague intent modifiers using sentiment analysis and word co-occurrence. Interactive text is displayed with the ability to adjust the ranges using slider widgets. (a) For a dataset of earthquakes in the US~\cite{usgs}, the system associates the vague modifier ``unsafe'' with the data attribute \texttt{magnitude}. Similar negative sentiment polarities (shown in \textbf{\red{red}}) result in a \texttt{top N} filter of magnitude $6$ and higher to be applied. (b) A dataset showing the health and wealth of nations~\cite{gapminder}. Here, the modifier ``struggling'' has a negative sentiment, while the \texttt{incomePerCapita} and \texttt{lifeExpectancy} attributes have positive sentiments shown in \textbf{\blue{blue}}. The diverging sentiment polarities result in \texttt{Bottom N} filters applied.}
 \label{fig:teaser}
}

\abstract{Natural language interaction with data visualization tools often involves the use of vague subjective modifiers in utterances such as ``show me the sectors that are \emph{performing}'' and ``where is a \emph{good} neighborhood to buy a house?.'' Interpreting these modifiers is often difficult for these tools because their meanings lack clear semantics and are in part defined by context and personal user preferences. This paper presents a system called \system that makes a first step in better understanding these vague predicates. The algorithm employs word co-occurrence and sentiment analysis to determine which data attributes and filters ranges to associate with the vague predicates. The provenance results from the algorithm are exposed to the user as interactive text that can be repaired and refined. We conduct a qualitative evaluation of the \system that indicates the usefulness of the interface as well as opportunities for better supporting subjective utterances in visual analysis tasks through natural language.} 


\CCScatlist{ 
 \CCScat{K.6.1}{Management of Computing and Information Systems}%
{Project and People Management}{Life Cycle};
 \CCScat{K.7.m}{The Computing Profession}{Miscellaneous}{Ethics}
}

\begin{document}
\maketitle


\section{Introduction\label{sec:introduction}}
 Understanding user intent in a query has been recognized as an important aspect of any natural language (NL) interaction system~\cite{li2010understanding,bos:2004}. Search queries typically consist of keywords and terms called \emph{modifiers} that imply a diverse set of search intents~\cite{jansen:2007}. While basic keyword matches from users' search queries might elicit a reasonable set of results, interpreting modifiers provides a better understanding of the semantics in the queries~\cite{manningbook}. 
 
Recently, NL interfaces for visual analysis tools have garnered interest in supporting expressive ways for users to interact with their data and see results expressed as visualizations~\cite{datatone,Setlur:2016,dhamdhere2017,ibmwatson,powerbi,thoughtspot,sun2010articulate,srinivasan2018orko,hoque2017applying}. Users often employ vague language while formulating natural language queries when exploring data such as ``which country has a \emph{high} number of gold medals?'' or ``what time of the day do \emph{more} bird strikes occur?''~\cite{hearst:2019}. There has been some precedence of research to better understand how these simple vague modifiers comprising of superlatives and numerical graded adjectives should be appropriately interpreted~\cite{setlur:iui,hearst:2019}. However, users also employ less concrete and often subjective modifiers such as `best', `safe', and `worse' in utterances~\cite{hearst:2019}. The interpretation of such modifiers makes it challenging for natural language interfaces to precisely determine the extensions of such concepts and mapping intent to the analytical functions provided in the visual analysis systems.

\noindent \textbf{Contribution}\\
This paper introduces \system,\footnote{The name \system is a portmanteau of `sentiment' and `modifier,' blending their concepts as they co-occur together.} a system to explore reasonable interpretations and defaults for such subjective vague modifiers in natural language interfaces for visual analysis. The algorithm identifies numerical attributes that can be associated with a modifier using word co-occurrence. Sentiment analysis determines the filter ranges applied to the attributes. Similar polarities result in associating the \texttt{Top N}  of data values for an attribute with the modifier, while diverging polarities are mapped to the \texttt{Bottom N}. 

Figure~\ref{fig:teaser}a indicates that `unsafe' and the attribute \texttt{magnitude} have similar negative sentiment polarities, defaulting to a higher earthquake magnitude range as seen in the map. The system has the ability to utilize any domain-specific information if available, such as WolframAlpha~\cite{wolframalpha}. Figure~\ref{fig:teaser}b shows diverging polarities for the modifier `struggling' paired with attributes \texttt{incomePerCapita} and \texttt{lifeExpectancy}. Lower numerical filter ranges based on the statistical properties of the data are applied to generate the scatterplot. Interactive text is displayed to show the provenance of the system's interpretation with clickable portions exposed as widgets that can be refined by the user. An evaluation of the system provides useful insights for future system design of NL input systems for supporting vague concepts in visual analysis.

\section{Related Work}
Research exploring the semantics of vague concepts for understanding intent transcends three main categories: (1) Computational Linguistics, (2) Intent and Modifiers in Search Systems, and (3) Natural Language Interaction for Visual Analysis.

\subsection{Computational Linguistics}
The notion of vagueness in language has been studied in the computational linguistics community~\cite{shapiro:2006}. Research has focused on the conceptualization and representation of vague knowledge~\cite{bobillo:2011}. The \textit{Vague} system introduces a technique for generating referring expressions that included gradable adjectives~\cite{deemter:2000}. De Melo et al. infer adjective grade ordering from large corpora~\cite{de-melo-bansal-2013-good} and Vegnaduzzo automatically detects subjective adjectives~\cite{Vegnaduzzo:2004}. Computational linguists have developed approaches for subjectivity and polarity labeling of word senses~\cite{weibe:2005,akkaya:2011}. In our work, we draw inspiration from linguistic literature, specifically polarity identification for computing the semantics around vague subjective concepts.

\subsection{Intent and Modifiers in Search Systems}
Search systems have explored techniques to deduce intent in queries during exploratory search. Several techniques exist to extract entity-oriented search intent to improve query suggestions and recommendations~\cite{duan:2015}. Detecting intent in search systems is also based on query topic classification~\cite{shen:2006}. Bendersky et al. assign weights to terms in a search query based on concept importance~\cite{bendersky:2010}. Recent work has focused on deriving query intent by fitting queries into templates~\cite{li2010understanding,agarwal:2010}.  Li et al. employ semantic and syntactic features to decompose queries into keywords and intent modifiers~\cite{li2010understanding}.  Researchers have predicted search intent and intentional task types from search behavior~\cite{cheng:2010,mitsui:2016}. While the goal of our work to interpret intent in queries is similar to that of search tasks, we focus on resolving vague modifiers to generate relevant visualization responses.
 
\subsection{Natural Language Interaction for Visual Analysis}
Similar to search systems, natural language interfaces for visual analysis need to understand intent and handle modifiers in the utterances. DataTone provides ambiguity widgets to allow a user to update the system's default interpretation~\cite{datatone}. Eviza and Analyza support simple pragmatics in analytical interaction through contextual inferencing~\cite{Setlur:2016,dhamdhere2017}. Evizeon~\cite{hoque2017applying} and Orko~\cite{srinivasan2018orko} extend pragmatics in analytical conversation. None of these systems consider how imprecise modifiers can be interpreted. The \textit{Ask Data} system describes the handling of numerical vague concepts such as `cheap' and `high' by inferring a range based on the underlying statistical properties of the data~\cite{setlur:iui}. Hearst et al. explore appropriate visualization responses to singular and plural superlatives and numerical graded adjectives based on the shape of the data distributions~\cite{hearst:2019}. We extend this work to more vague, subjective modifiers.

\section{The \system System}
We introduce a system, \system that interprets vague modifiers such as `safe' and `struggling' in a NL interface for visual analysis. The system employs a web-based architecture with the input query processed by an ANTLR parser with a context-free grammar, similar to parsers described in~\cite{Setlur:2016,hoque2017applying}. A data manager provides information about the data attributes and executes queries to retrieve data. The query upon execution, generates a D3 visualization result~\cite{d3}.

\subsection{Algorithm for Interpreting Vague Modifiers}
The process for resolving a set of data attributes and their values to a modifier found in the NL input to \system, is outlined as: 
\vspace{-4mm}
\begin{algorithm}
\KwIn{Natural language utterance $\alpha$}
\KwOut{Generate visualization response}
\nonl $\alpha$ is the NL input utterance.\\
\nonl $m$ is the vague modifier in the utterance $\alpha$.\\
\nonl Part-of-Speech tagger $POS$ identifies $m$ in $\alpha$.\\
\nonl $attrs_{num}$ is the set of numerical attributes in the dataset $D$.\\
\nonl $attrs_{cnum}$ is the set of co-occurring numerical attributes in $D$ with $attrs_{cnum} \in attrs_{num}$.\\
\nonl $PMI$ computes co-occurrence scores $w_{c}$ for $m$ and $attrs_{num}$.\\
\nonl $polarity$ computes sentiment polarities $p$ for $m$ and $attrs_{cnum}$.\\
Invoke $POS(\alpha)$ returning $m$.\\
Compute $PMI(m, attrs_{num}) \rightarrow$ $w_c$ for each $attr_i \in attrs_{cnum}$.\\
Compute $polarity(m, attrs_{cnum}) \rightarrow$ $p$.\\
Update interface based on $w_c$ and $p$.
\caption{Interpretation of Vague Modifiers in \system}
\label{algo:sentifiers}
\end{algorithm}
\vspace{-4mm}

\subsubsection{Parse Vague Modifiers}
Vague modifiers are gradable adjectives that modify nouns and and are associated with an abstract scale ordered by their semantic intensity~\cite{kennedy:1997}. Gradable adjectives can be classified into two categories based on their interpretation as measure functions~\cite{kennedy:1997}. Numerical graded adjectives such as `large' and `cheap' are viewed as measurements that are associated with a numerical quantity for size and cost respectively.  Complex graded adjectives like `good' and `healthy' tend to be underspecified for the exact feature being measured. 

While the interpretation of numerical gradable adjectives has been explored in NL interfaces for visual analysis~\cite{Setlur:2016,setlur:iui,hearst:2019}, this paper specifically focuses on the handling of complex gradable adjectives. \system first applies a commonly used performant part-of-speech (POS) tagger during the parsing process to identify these complex gradable adjectives and their referring attributes in the NL utterances~\cite{Toutanova:2003}. The system can distinguish complex gradable adjectives by checking for the absence of superlative or comparative tags that are used to annotate numerical graded adjectives.

\subsubsection{Compute Modifier and Attribute Co-occurrence Scores}
\begin{figure}[ht]
\centering
\includegraphics[width=.8\columnwidth]{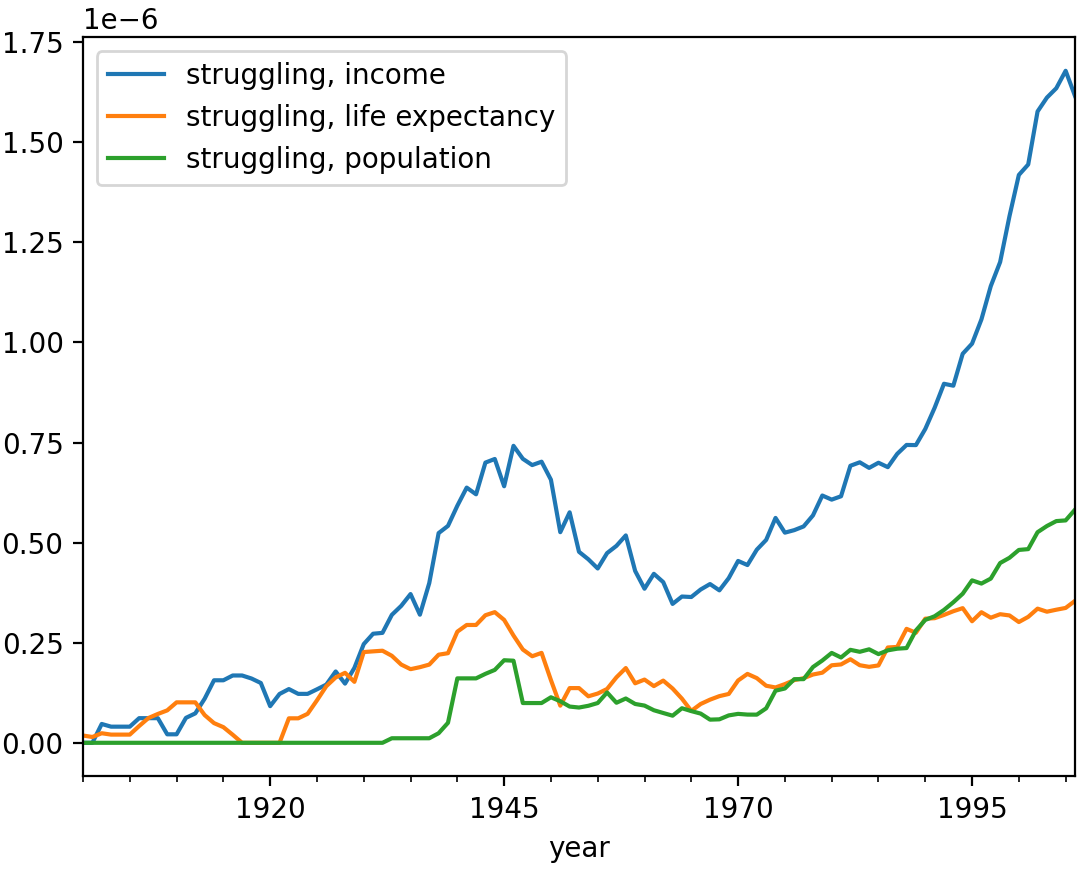}
\caption{PMI values for the modifier `struggling' with each of the attribute n-grams, `income', `life expectancy', and `population' in the Google n-gram corpus. Higher PMI scores indicate a higher co-occurrence of the modifier and attribute terms.}
\label{fig:cooccurrence}
\vspace{-5mm}
\end{figure}

The next step maps the vague modifier to a scale based on its semantic intensity so that the modifier can be interpreted as a set of numerical filters for generating a visualization response. We base our approach on linguistic models that represent the subjectivity of complex modifiers as a generalized measure mapping the modifier to numerical attributes in a multidimensional space~\cite{galit:2016}. For example, the subjectivity of the modifier `healthy' can be interpreted based on `weight', `amount of exercise', and `hospital visits.'  

\system computes the semantic relatedness between the modifier and the numerical data attributes using a co-occurrence measure. To have sufficient coverage for co-occurrence, we use an extensive Google n-grams\footnote{An n-gram is a contiguous sequence of n items from a sequence of text.} corpus~\cite{Michel176}. To maximize the chances of co-occurrence, \system considers co-occurrence between all n-gram combinations of the modifier and the attribute names. For example, some of the n-grams for the attribute \texttt{income per capita} are `income per capita,' `income per,' `per capita,' and `income.' 

We employ a Pointwise Mutual Information Measure (PMI), an information-theoretic measure that quantifies the probability of how tightly occurring a modifier $m$ and a numerical attribute $attr_{num}$ are to the probability of observing the terms independently~\cite{church-hanks-1989-word}. We found this measure to work well and was performant with terse word co-occurrence pairings without requiring sentence embeddings. We consider any numerical attribute $attr_{cnum}$ that has a non-zero PMI score, indicating the presence of a co-occurrence with $m$. The PMI of modifier n-gram $t_m$ with one of the attribute n-grams $t_{attr}$ is:

\setlength{\belowdisplayskip}{-5pt} \setlength{\belowdisplayshortskip}{-5pt}
\setlength{\abovedisplayskip}{-5pt} \setlength{\abovedisplayshortskip}{-5pt}

\begin{equation}
PMI(t_m,t_{attr}) = log \frac{p(t_m,t_{attr})}{p(t_m)p(t_{attr})}
\end{equation}

\subsubsection{Determine Sentiment Polarities}
Once the modifier is semantically associated with co-occurring numerical attributes, we need to determine a reasonable numerical range to associate with the modifier. Sentiment polarity analysis is a linguistic technique that uses positive and negative lexicons to determine the polarity of a phrase~\cite{wilson-etal-2009-articles}. The technique provides the ability to dynamically compute the sentiment of the phrase based on the context in which its terms co-occur rather than pre-tagging the phrase with absolute polarities, which is often not scalable.

\begin{figure}[ht]
	\centering \includegraphics[width=.89\columnwidth]{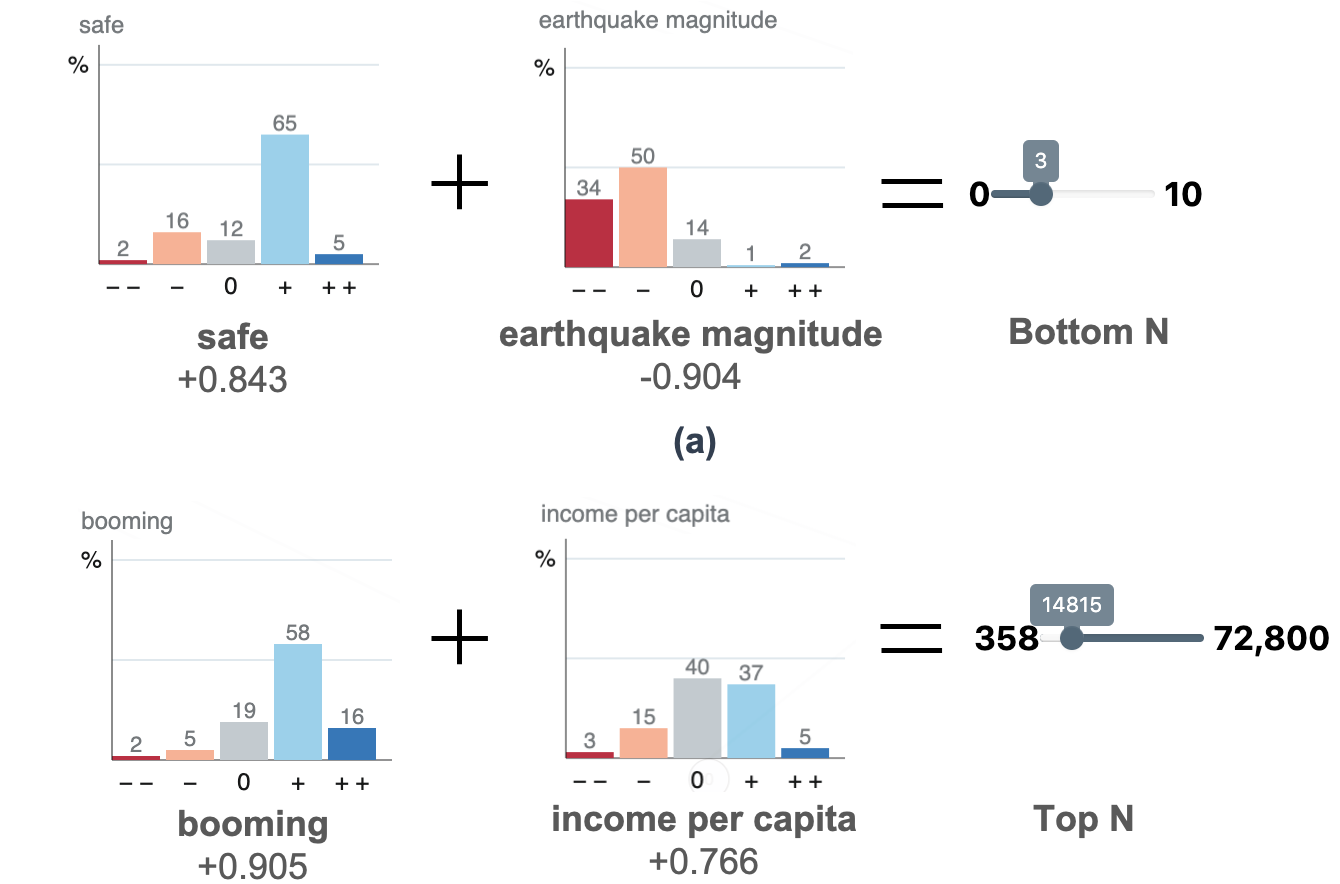}
	\caption{Sentiment polarity logic with sentiments and their normalized scores for the modifiers and the numerical attributes. (a) The modifier `safe' and attribute \texttt{earthquake\_magnitude} have positive and negative sentiments respectively, resulting in a \texttt{Bottom N} range based on the Richter scale~\cite{wolframalpha}. (b) The modifier `booming' and attribute \texttt{income per capita} both have positive sentiments, resulting in a \texttt{Top N} computed based on statistical data properties.}
	\label{fig:polarity}
	\vspace{-6mm}
\end{figure}

We determine the individual sentiment scores with a sentiment classification based on a recursive neural tensor network~\cite{socher-etal-2013-recursive}. We choose this technique as its models handle negations and reasonably predict sentiments of terser phrases, characteristic of queries to \system. The sentiments are returned as a 5-class classification: very negative, negative, neutral, positive, and very positive. The values are normalized as $[-1, +1]$, ranging from negative to positive to provide an overall sentiment. We then determine the sentiment polarities of the modifier $m$ and co-occurring attribute $attr_{cnum}$ pair based on their individual sentiments (ignoring the strength of the sentiments) using the following combinatorial logic. We treat neutral sentiment similar to positive sentiment as neutral text tends to lie near the positive boundary of a positive-negative binary classifier~\cite{wilson-etal-2009-articles}.

\begin{algorithmic}
  \IF {($sentiment_{m} == positive$ \OR $sentiment_{m} == neutral$) \AND ($sentiment_{attr_{cnum}} == positive$ \OR $sentiment_{attr_{cnum}} == neutral$)}
 \STATE Compute $TopN(attr_{cnum})$.
  \ELSIF {($sentiment_{m} == positive$ \OR $sentiment_{m} == neutral$) \AND $sentiment_{attr_{cnum}} ==negative$}
 \STATE Compute $BottomN(attr_{cnum})$.
  \ELSIF {$sentiment_{m} == negative$ \AND ($sentiment_{attr_{cnum}} == positive$ \OR $sentiment_{attr_{cnum}} == neutral$)}
 \STATE Compute $BottomN(attr_{cnum})$.
 \ELSIF {$sentiment_{m} == negative$ \AND $sentiment_{attr_{cnum}} ==negative$}
 \STATE Compute $TopN(attr_{cnum})$.
  \ENDIF
\end{algorithmic}

\begin{figure*}[ht]
	\centering 
	\includegraphics[width=.8\textwidth]{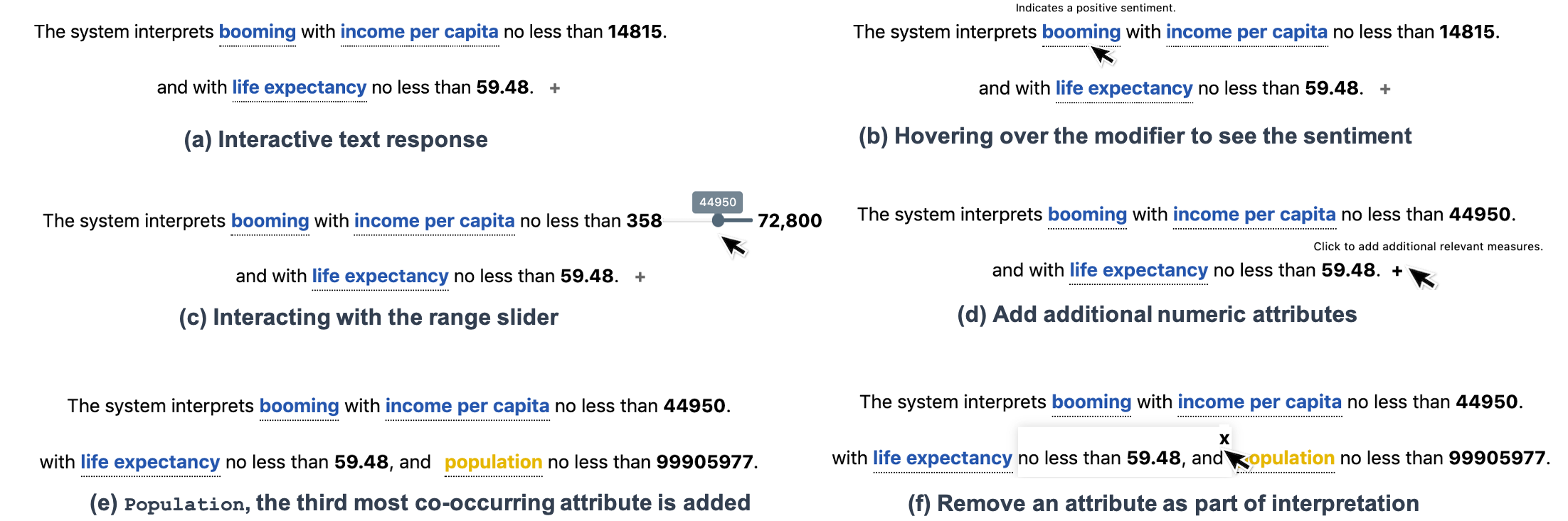}
	\caption{Interactive text response to a query ``which countries are \emph{booming}?''. \system provides the ability to refine the system defaults.}
	\label{fig:interface}
	\vspace{-4mm}
\end{figure*}

\system uses sentiment polarities to compute the ranges in two ways: If domain knowledge exists, the system uses the information to determine a default (Figure~\ref{fig:polarity}a uses the Richter scale~\cite{wolframalpha}). Otherwise, the system computes \texttt{Top N} to range from $[med + MAD,max]$ and \texttt{Bottom N} to range from $[min, abs(med - MAD)]$ where $med$, $MAD$, $min$, and $max$  are the median, median absolute deviation, minimum, and maximum values for $attr_{cnum}$ respectively (see Figure~\ref{fig:polarity}b). We choose $MAD$ as it tends to be less affected by non-normality~\cite{cairns_2019}. 
\vspace{-2mm}

\subsection{User Interface}
Figure~\ref{fig:teaser} shows the \system interface with an input field that accepts text queries. Upon execution of the query, range filters for the co-occurring numerical attributes are applied, showing a visualization response. The system interpretation is expressed in the form of interactive text~\cite{bretvictor} above the visualization (Figure~\ref{fig:interface}a) to help the user understand the provenance of how the modifier was interpreted. 

Positive, negative, and neutral sentiments are shown in \textbf{\blue{blue}}, \textbf{\red{red}}, and \textbf{\yellow{yellow}}  respectively (Figure~\ref{fig:interface}b). The text contains widgets that show ranges starting from the highest co-occurring one. Similar to other NL systems~\cite{datatone,Setlur:2016,setlur:iui}, we expose system presumptions as widgets (Figure~\ref{fig:interface}c). If domain-specific semantics are used, a link to the source is provided (Figure~\ref{fig:teaser}a). To provide easier readability, \system displays up to two widgets. Word co-occurrence and sentiment analysis techniques can result in incorrect results. The user has the ability repair the system decisions (Figures~\ref{fig:interface}d and f) and the interface updates to reflect the changes (Figure~\ref{fig:interface}e). These refinements are persistent for the duration of the user session.

\section{Evaluation}
We conducted a user study of \system with the following goals: (1) collect qualitative feedback on the handling of the modifiers for various visual analysis tasks and (2) identify system limitations. This information would help provide insights as to how the handling of complex vague modifiers could integrate into a more comprehensive NL visual analysis interface. The study was exploratory in nature where we observed the types of vague modifiers people asked and how they responded to the system behavior. Because the main goal of our study was to gain qualitative insight in the system behavior, we encouraged participants to think aloud with the experimenter.

\subsection{Method}
\vspace{-.5mm}
\subsubsection{Participants}
We recruited ten volunteers (five males, five females, age 24 -- 65). All were fluent in English and all regularly used some type of NL interface such as Google. Eight used a visualization tool on a regular basis and the rest considered themselves beginners.

\subsubsection{Procedure and Apparatus}
Each participant was randomly assigned a dataset of earthquakes in the US~\cite{usgs} or the health and wealth of nations~\cite{gapminder} with equal number of participants for each. We began with a short introduction of how to use the system. Participants were instructed to phrase their queries in whatever way that felt most natural and to tell us whenever the system did something unexpected. We discussed reactions to system behavior throughout the session and concluded with an interview. The study trials were done remotely over a shared videoconference to conform with social distancing protocol due to COVID-19. All sessions took approximately 30 minutes. 

\subsubsection{Analysis Approach}
We employed a mixed-methods approach involving qualitative and quantitative analysis, but considered the quantitative analysis mainly as a complement to our qualitative findings. The quantitative analysis consisted of the number of times participants used vague subjective modifiers and interacted with the text response. 

\subsection{Results and Discussion}
Overall, participants were positive about the system and identified many benefits. Several participants were impressed with the ability of the system to understand their queries (``I typed \emph{scary} to see what it would do, and it understood.'' [$P2$]). \system' text feedback was found to be helpful (``I wasn't sure how the system would handle this, but it was pretty clear when I saw the response'' [$P4$]). The participants appreciated the functionality to be able to correct the system's response (``I wanted to tweak the range a bit and it was useful to be able to change the slider and see the result update'' [$P9$]). 

The number of unique vague modifiers per participant ranged from $3$ to $12$ ($\mu = 6.7$) with a total of $24$ unique complex modifiers overall. The three most common modifiers were `good', `bad', `severe' for the earthquakes dataset and `prosperous', `flourishing', `poor` for the health and wealth of nations dataset. All participants interacted with the text response to understand the system behavior. The most common interaction was updating the data ranges for the attributes (69\% of the interactions), followed by adding new attributes (23\%), and deleting attributes from the interpreted result (8\%). Comments relevant to this behavior included, ``The range seemed high for me and I changed it. It was nice to see the system remember that'' [$P10$], ``I wanted population to be added to the mix and it was easy to just click and do that'' [$P3$], and ``I wasn't interested in life expectancy so I just got rid of it'' [$P1$]. 

The study also revealed several shortcomings and provides opportunities for future NL systems supporting visual analysis tasks:

\noindent \textbf{Support for more complex interpretations:} The current implementation does not support combinations of vague modifiers in the same query. For example, the system was unable to interpret ``show me countries that are \emph{doing very well} and \emph{poorly}.'' [$P4$].  $P2$ expressed that they wanted flexibility in defining analytical functions such as associating `unsafe` to the frequency of recently occurring earthquakes with magnitude $5$ are greater. \system failed to correctly interpret queries such as ``which countries are \emph{reasonably doing well},'' where $P7$ expected some middle range, though they were able to adjust the ranges after.
A comprehensive evaluation with additional datasets would be necessary to ascertain how effective this system would be alongside standard visual analysis tools.

\noindent \textbf{Handling customization and in-situ curation:}  The topic of customization of the interpretation behavior came up during the study. For example, $P1$ said ``I typed - show me which countries are \emph{affordable} and it showed me an income range. I was expecting a response that considered inflation, GDP or have a way for me to define that.'' The algorithms employed in \system assume that the data attributes are curated with human-readable words and phrases. However, data is often messy with domain-specific terminology. Future work should explore mechanisms for users to customize semantics of attributes and interpretations in the flow of their analysis.

\noindent \textbf{Handling system expectations, biases, and failures:} 
NL algorithms have shown to exhibit socio-economic biases, including gender and racial assumptions often due to the nature of the training data~\cite{bias}. Their use can perpetuate and even amplify cultural stereotypes in NL systems. For example, $P7$ commented, ``I asked for \emph{good} places to live and the system responded with a high income per capita. To me, that opens up bigger issues such as gentrification and economic segregation.'' This suggests that there is a responsibility for improved transparency in system behavior; determining appropriate de-biasing methods remains an open research problem.

\section{Conclusion}
This paper presents a technique to explore how a system can interpret subjective modifiers prevalent in natural language queries during visual analysis. Using word co-occurrence and sentiment polarities, we implement \system to map these modifiers to more concrete functions. We expose the provenance of the system's behavior as an interactive text response. An evaluation of the system indicates that participants found the system to be intuitive and appreciated the ability to refine the system choices. Feedback from interacting with \system identifies opportunities for handling vagueness in language in the future design of such natural language tools to support data exploration. As Bertrand Russell stated~\cite{russell1921analysis} -- ``Everything is vague to a degree you do not realize till you have tried to make it precise."

\newpage
\bibliographystyle{abbrv-doi}

\bibliography{sentifiers}
\end{document}